\documentclass[12pt,twocolumn]{article}
\pdfoutput=1
\usepackage[utf8]{inputenc}
\usepackage{authblk}
\usepackage{siunitx}
\usepackage{amsmath}
\usepackage{graphicx}
\usepackage[numbers]{natbib}
\usepackage{hyperref}
\usepackage{subcaption}
\everymath{\bf}
\everydisplay{\bf}
\title{Algorithms for Least-Squares Noncartesian MR Image Reconstruction}
\author{Tobias C Wood \\\href{mailto:tobias.wood@kcl.ac.uk}{tobias.wood@kcl.ac.uk}}
\affil{Department of Neuroimaging, King's College London}
\date{}

\begin{document}

\maketitle

\section*{Abstract}

Iterative least-squares MR reconstructions typically use the Conjugate Gradient algorithm, despite known numerical issues. This paper demonstrates that the more recent LSMR algorithm has favourable numerical properties, and is to be preferred in situations where Toeplitz embedding cannot be used to accelerate the Conjugate Gradient method.

\section*{Introduction}

The seminal paper of Pruessman et al\cite{Pruessmann:2001} introduced the concept of formulating MR image reconstruction as a least squares problem:

\begin{equation}
    Ax=b + \mathnormal{\epsilon}
\label{eq1}
\end{equation}

where $x$ represents the desired image as a vector of complex-valued voxel values, $b$ is a similar vector containing the acquired k-space samples, $\epsilon$ is complex-valued Gaussian-distributed noise, and $A$ is the system matrix (often also denoted as the encoding operator/matrix $E$ in the MR literature). This represents all the steps involved in the MR image formation process, which generally include splitting the image into multiple receiver channels, Fourier transformation, and sampling along the desired trajectory.

In general for MRI the matrix $A$ is too large to store in memory or efficiently invert, which precludes using the closed-form solution to~\eqref{eq1}. Instead Pruessman et al. expressed the sensitivity, Fourier, and sampling steps as linear operators and solved~\eqref{eq1} using the iterative Conjugate Gradient (CG) algorithm~\cite{Pruessmann:2001}. This has become popular enough that the term ``cgSENSE'' has entered the MRI lexicon~\cite{Maier:2020}. CG is incredibly simple to implement, features superior convergence performance compared to simple gradient descent and Toeplitz embedding can be used to avoid repeated gridding operations which leads to a significant computation speed increase for noncartesian trajectories~\cite{Wajer:2001, Fessler:2005, Baron:2018}.

However, CG has some numerical properties that are not well suited to MR reconstruction~\cite{Pruessmann:2001}. These problems arise because, in general, the system matrix $A$ for MRI is not square or symmetric (due to any or all of multiple channels, undersampling, and a noncartesian trajectory). CG requires a positive definite system matrix, and to guarantee this instead of directly solving~\eqref{eq1} we solve the normal equations

\begin{equation}
    A^{\dagger}Ax=A^{\dagger}b
\label{eqnormal}
\end{equation}

where $^{\dagger}$ denotes the adjoint (conjugate transpose). This squares the condition number of the system, which makes it more sensitive to random noise and leads to very slow convergence of 3D noncartesian trajectories (which have a large condition number due to the repeated sampling of the center of k-space). In addition preconditioning, a standard method to reduce the effective condition number, can only be used in image space and not k-space~\cite{Ong:2020}. Non-iterative noncartesian reconstructions use Sample Density Compensation (SDC) weightings in k-space to reconstruct meaningful images~\cite{Pipe:2000, Johnson:2009, Zwart:2012}. While SDC weightings can be, and often are, inserted in an ad-hoc way into cgSENSE, this results in noise inflation~\cite{Pruessmann:2001, Baron:2018}.

My particular motivation is reconstructing Zero Echo-Time (ZTE) images, which necessitate a 3D radial center-out trajectory for which the above issues are particularly acute~\cite{Ljungberg:2021}. ZTE can capture signals from protons with short transverse relaxation times which are normally MR invisible~\cite{Weiger:2019, Weiger:2020, Baadsvik:2021}. A standard clinical MRI setup contains foam cushioning and structural plastic in the receive coil that contain significant amounts of such protons, and so contain signals outside the nominal field of view. This enforces an enlarged reconstruction field of view to avoid aliasing. Toeplitz embedding would require an additional eight-fold memory requirement (two times in each dimension)~\cite{Baron:2018}. For reasonably sized images, for example 1\si{\milli\meter} isotropic or better full brain images, the large memory requirements this imposes quickly become impractical. This limitation motivated a search for alternative algorithms.

Two such algorithms are LSQR~\cite{LSQR} and the much more recent LSMR~\cite{LSMR}\footnote{The original papers for LSQR and LSMR do not define the names as acronyms. However I believe they stand for ``Least-Squares via QR factorization'' and ``Least-Squares Minimum Residual'' respectively}. These both use a Golub-Kahan bi-diagonolization process, which is computationally similar to CG. To quote the Stanford University Systems Optimization Laboratory website: ``[LSQR] is algebraically equivalent to applying CG to the normal equation but has better numerical properties, especially if $A$ is ill-conditioned'', and ``LSQR reduces $|r|$ monotonically (where $r = b - Ax$ ... ) On least-squares problems, if an approximate solution is acceptable (stopping tolerances quite large), LSMR may be a preferable solver because it reduces both $|r|$ and $|A'r|$ monotonically and may be able to terminate significantly earlier''. Because both algorithms conduct operations in image space and k-space they support preconditioning in either domain~\cite{Orban:2020}. Ong et al recently derived a k-space preconditioner that gave superior results to SDC in the context of regularized reconstructions using the Primal-Dual Hybrid Gradient (PDHG) algorithm~\cite{Ong:2020}. Both algorithms support simple Tikhonov regularization through a simple algorithmic alteration.

In this paper I demonstrate that, when Toeplitz embedding is impractical, LSQR gives comparable performance to CG while LSMR outperforms both. LSMR hence may be of interest to the wider MR community for reconstructing large noncartesian MR images.

\section*{Methods}

I compared the performance of the CG, LSQR and LSMR algorithms using two different datasets: the 2D radial dataset from the 2020 ISMRM Reproducibility Challenge~\cite{Maier:2020} and an in-vivo 3D radial ZTE scan. Acquisition details for the 2D dataset are given in the reference. The 3D dataset was acquired on a 3T scanner equipped with a 48-channel head coil (GE Healthcare) at 1\si{\milli\meter} isotropic resolution, 220 isotropic matrix size, flip-angle $\mathrm{2\si{\degree}}$, $\mathrm{TR/TI/T_{recovery}=2/800/600\si{\milli\second}}$, 2x radial oversampling, $\mathrm{2\pi}$ spoke undersampling, 512 spokes per segment and a phyllotaxis trajectory~\cite{Emil:2022}. 

All reconstructions were carried out with the RIESLING toolbox~\cite{Wood:2021}. The Jupyter notebooks used for the analysis are available at \href{https://github.com/spinicst/riesling-algorithms}{https://github.com/spinicst/riesling-algorithms}. Reconstructions used Apple MacBook Pro with an M1 Pro chip and 32 gigabytes of RAM. The 3D data was first compressed from to 12 virtual channels~\cite{Huang:2008}, and then the same steps were followed for both 2D and 3D. Sensitivity maps were extracted directly from the fully-sampled central k-space region~\cite{Yeh:2005}. For CG, SDC weights were calculating with Pipe's method~\cite{Zwart:2012} while for LSQR and LSMR I used Ong's single-channel k-space preconditioner~\cite{Ong:2020}. Both datasets were then reconstructed twice with each algorithm, once without SDC/preconditioning as appropriate and once without, yielding six different reconstructions for each dataset. All algorithms were terminated at a fixed limit of 32 iterations. The results were compared both at this final iteration, and earlier at a somewhat arbitrary choice of the sixth iteration for the 2D dataset and the 4th iteration for 3D.

\section*{Results}

\begin{figure*}[ht]
\begin{center}
     \begin{subfigure}[b]{0.5\textwidth}
         \includegraphics[width=\textwidth]{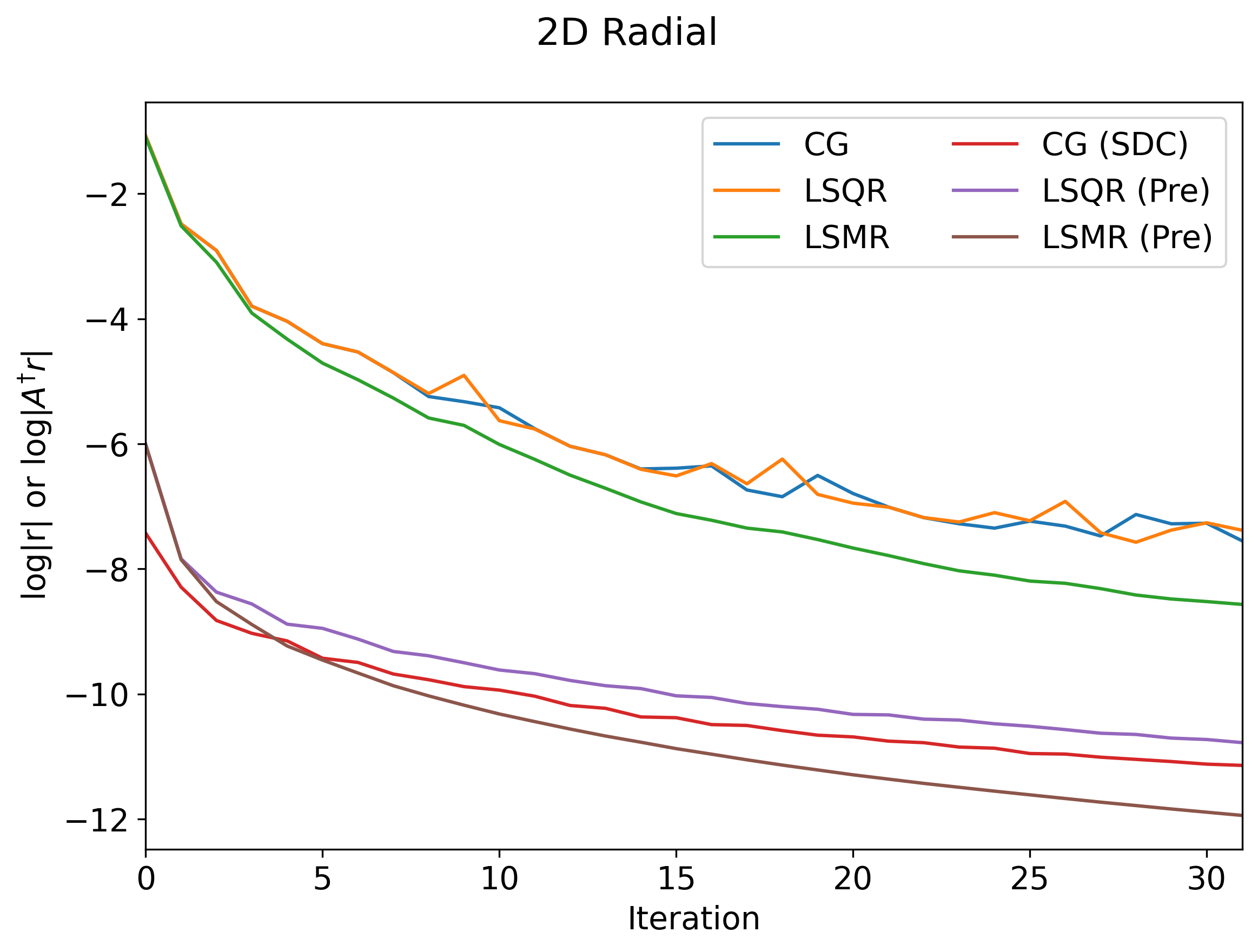}
     \end{subfigure}%
     \begin{subfigure}[b]{0.5\textwidth}
         \includegraphics[width=\textwidth]{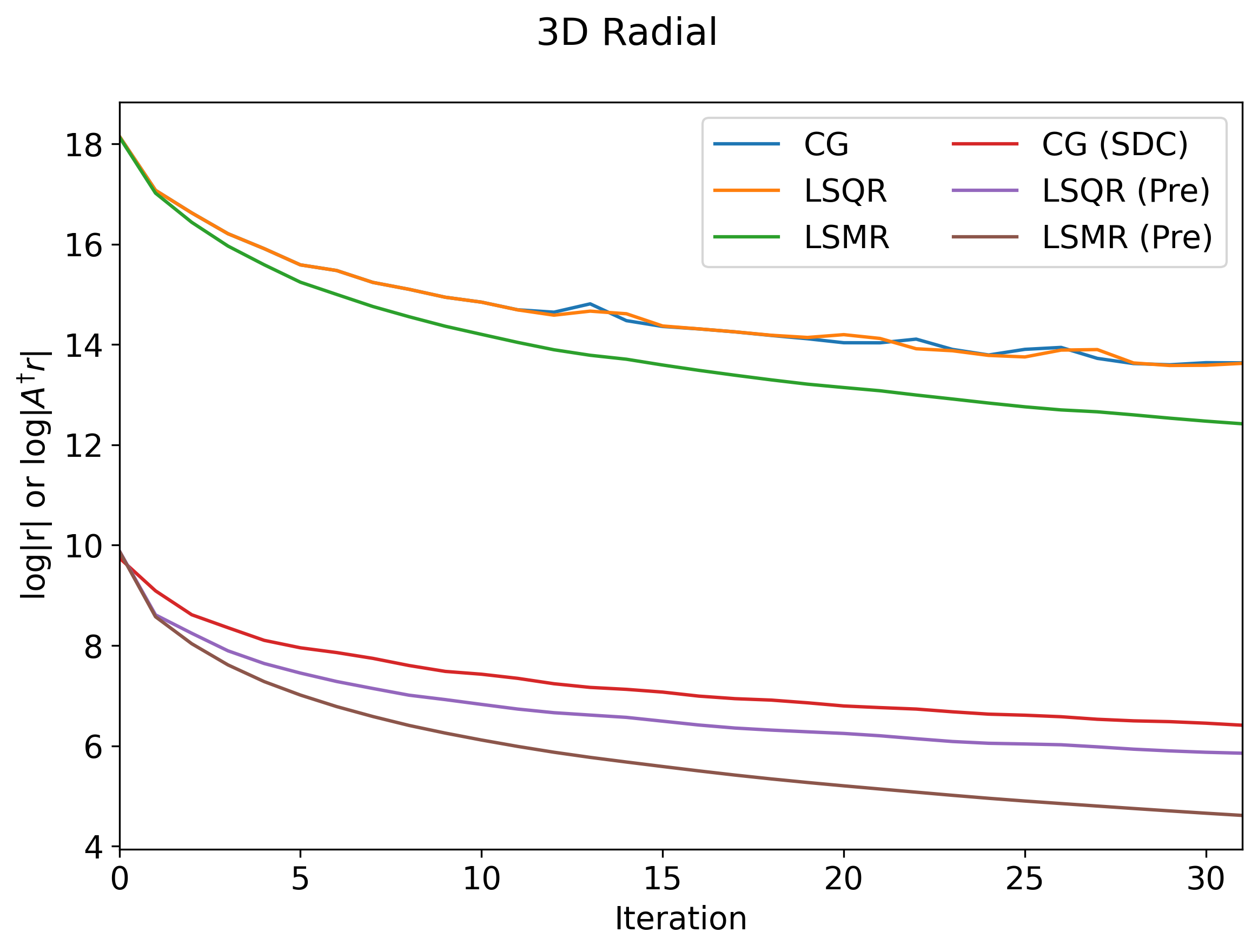}
     \end{subfigure}%
\caption{The logarithm of the norm of the image-space residual for the 2D (left) and 3D (right) datasets for all the algorithms investigated.}
\label{fig-resid}
\end{center}
\end{figure*}

\begin{figure*}[htb]
\begin{center}
     \begin{subfigure}[b]{0.5\textwidth}
         \includegraphics[width=\textwidth]{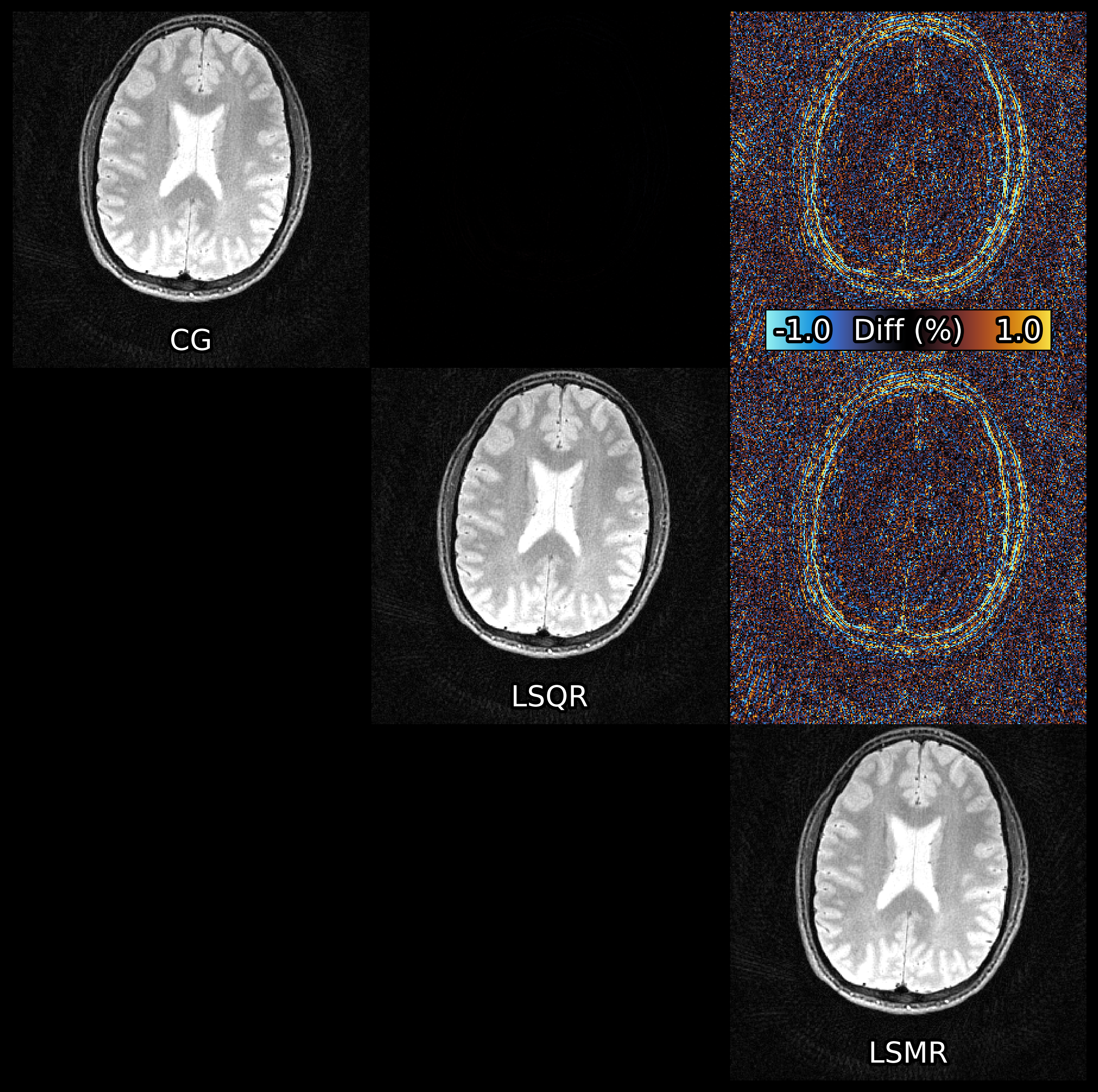}
     \end{subfigure}%
     \begin{subfigure}[b]{0.5\textwidth}
         \includegraphics[width=\textwidth]{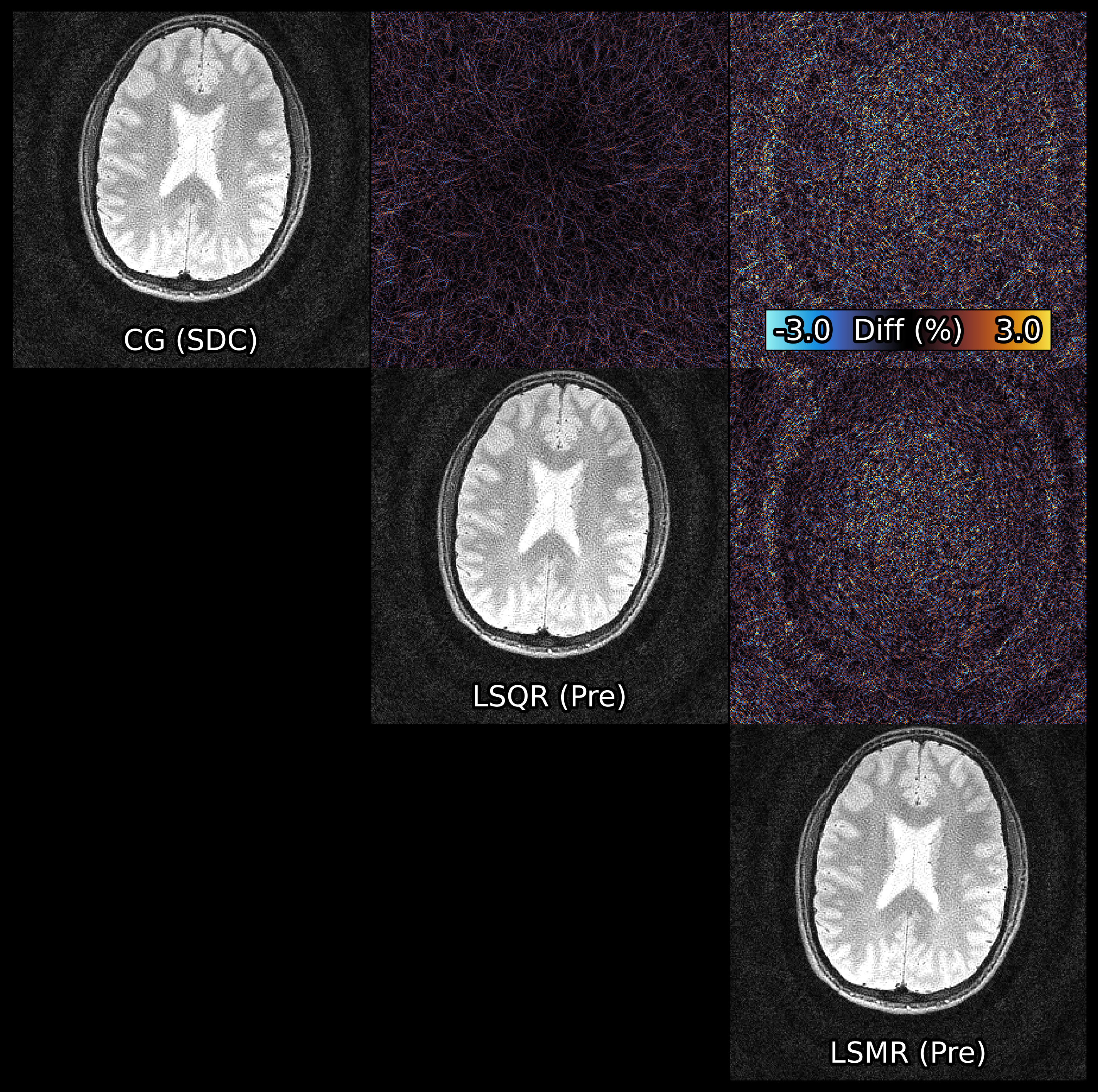}
     \end{subfigure}%
\caption{Reconstructed images at the final iteration for the 2D data without (left) and with (right) SDC or preconditioning. The images are shown on the diagonal and percent (normalised to the highest intensity value in the CG image) differences between them on the upper off-diagonals. }
\label{2d-final}
\end{center}
\end{figure*}

\begin{figure*}[htb]
\begin{center}
     \begin{subfigure}[b]{0.5\textwidth}
         \includegraphics[width=\textwidth]{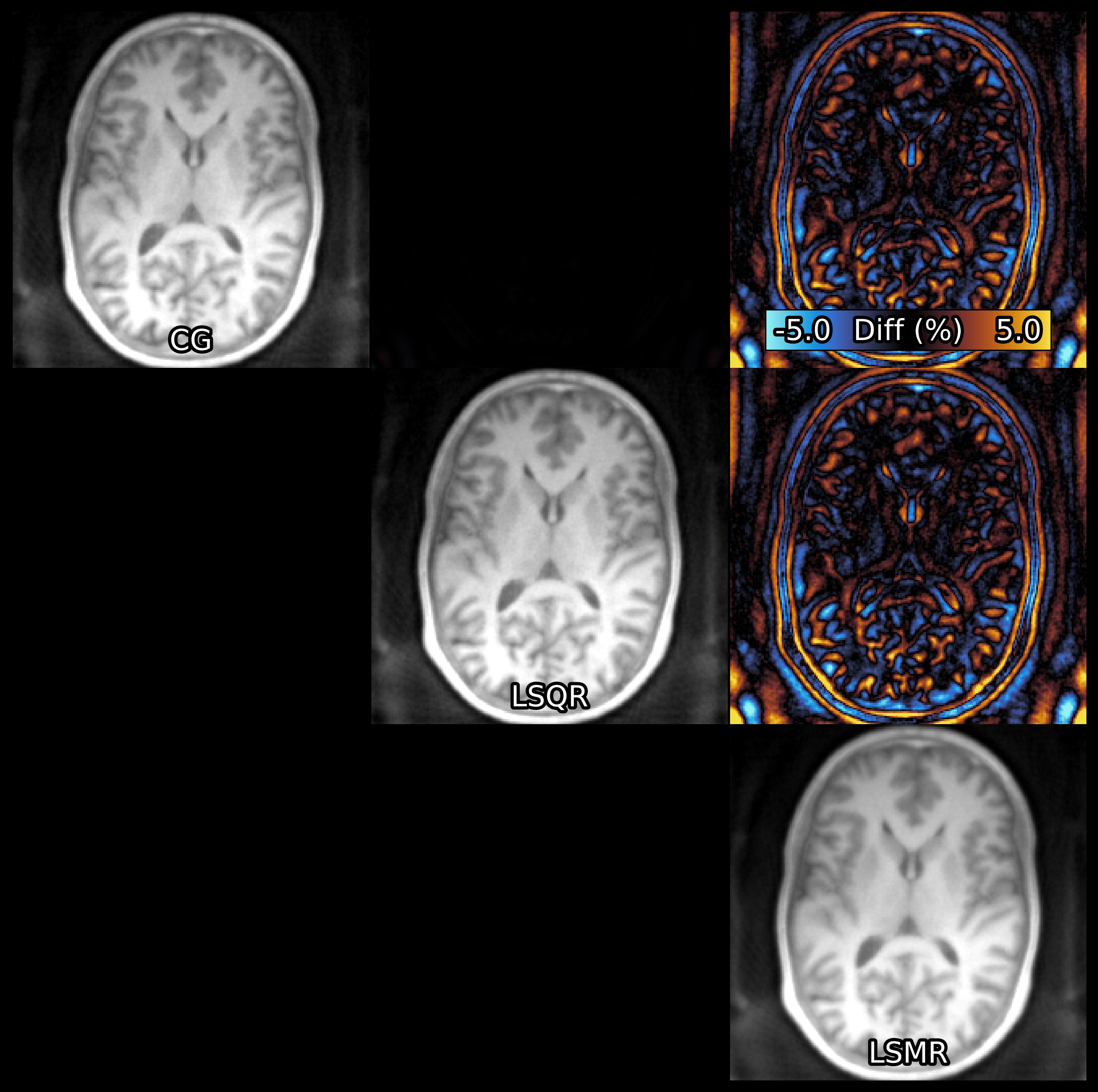}
     \end{subfigure}%
     \begin{subfigure}[b]{0.5\textwidth}
         \includegraphics[width=\textwidth]{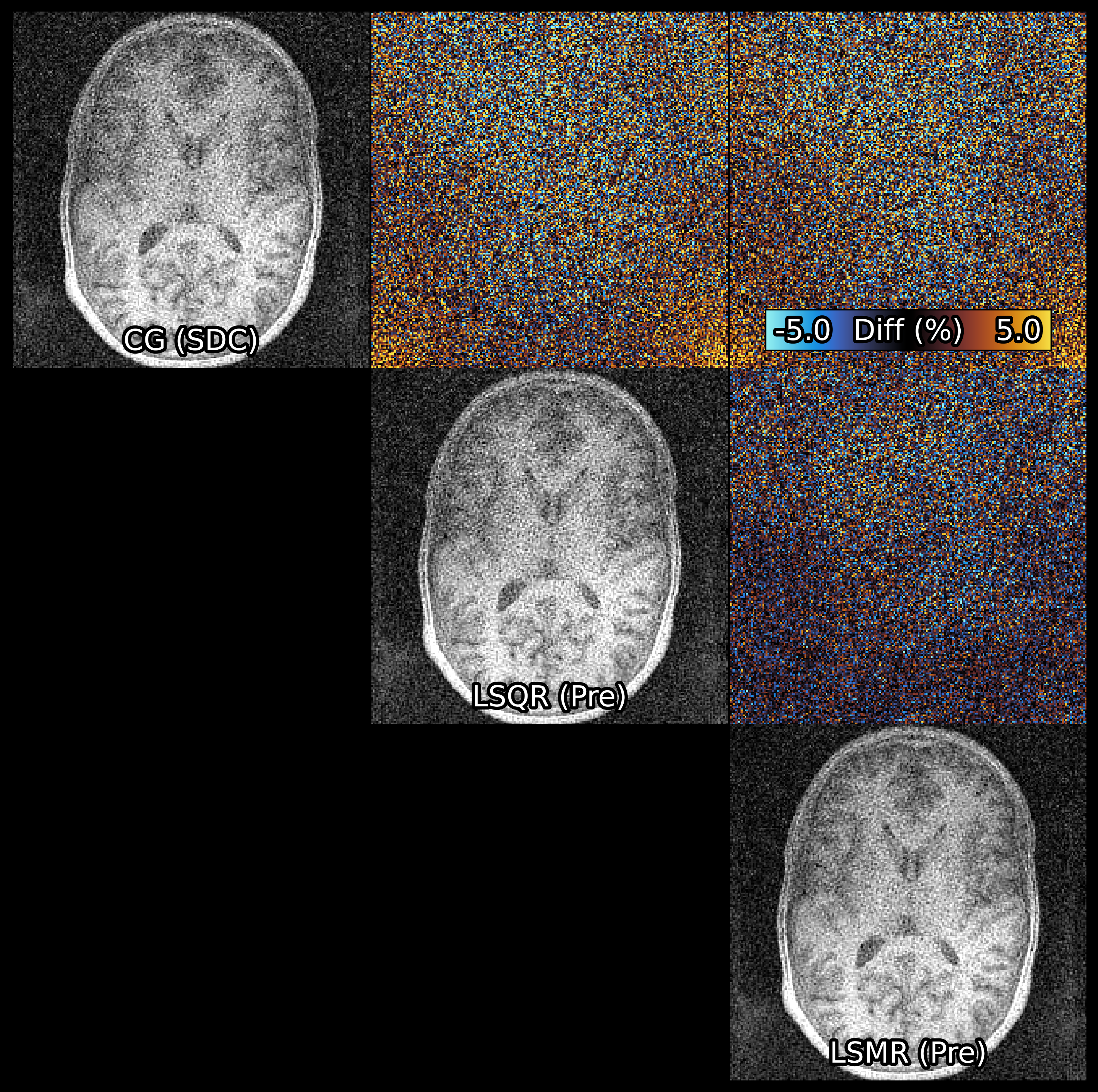}
     \end{subfigure}%
\caption{Reconstructed images at the final iteration for the 3D data without (left) and with (right) SDC or preconditioning. The images are shown on the diagonal and percent (normalised to the highest intensity value in the CG image) differences between them on the upper off-diagonals. }
\label{3d-final}
\end{center}
\end{figure*}

Figure 1 shows the convergence behaviour for all algorithms and both datasets. The quantity of interest is the norm of the image-space residual, which is the $|A^{\dagger}r|$ term for LSQR and LSMR as the primary residual is in k-space. CG and LSQR produce near identical results without SDC/preconditioing, including oscillatory behaviour with iterations where the residual temporarily increases. The addition of SDC/preconditioning reduces the residual norm for all three algorithms, and leads to differences in the residual between CG and LSQR. LSMR produces a smaller final residual in all cases after a few iterations, although in the 2D case CG does outperform it initially. The computation of all three algorithms was approximately the same, at 7 iterations per second for 2D and 1 iteration per 30 seconds for 3D (including the time taken to save per-iteration data to generate the figures in this paper).

Figures 2\&3 show the reconstructed images at the final iteration for each algorithm and the differences between them. Matching the residual results, without SDC/preconditioning, CG and LSQR produce essentially identical images. The 2D reconstructions are crisp and high quality, while the 3D reconstructions are blurred. Significant signal can be observed outside the head from cushioning in the 3D images due to the ZTE acquisition. The images from LSMR are nearly indistinguishable to CG and LSQR, but in the 2D case small but elevated differences are evident on the edges of the skull while for the 3D data much larger differences are apparent, including within the brain on grey matter/white matter boundaries.

The addition of SDC/preconditioning leads to differences between CG and LSQR, which perhaps show some structure in the 2D case but are only noise in the 3D case. There are no anatomical features in the difference between LSMR and the other two algorithms, but LSMR arguably displays a lower background noise level. The 3D images with SDC/preconditioning appear very noisy.

\begin{figure*}[htb]
\begin{center}
     \begin{subfigure}[b]{0.5\textwidth}
         \includegraphics[width=\textwidth]{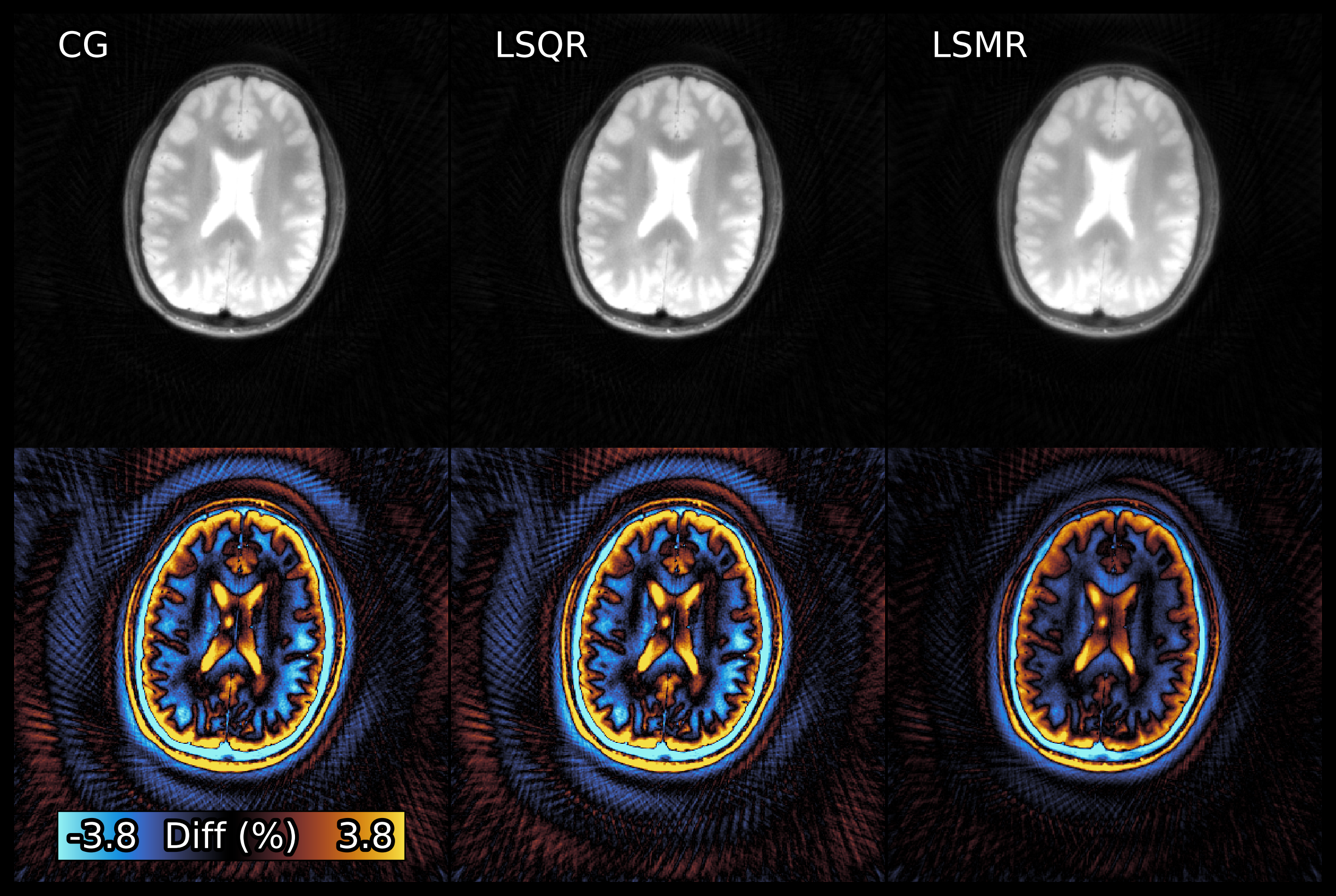}
     \end{subfigure}%
     \begin{subfigure}[b]{0.5\textwidth}
         \includegraphics[width=\textwidth]{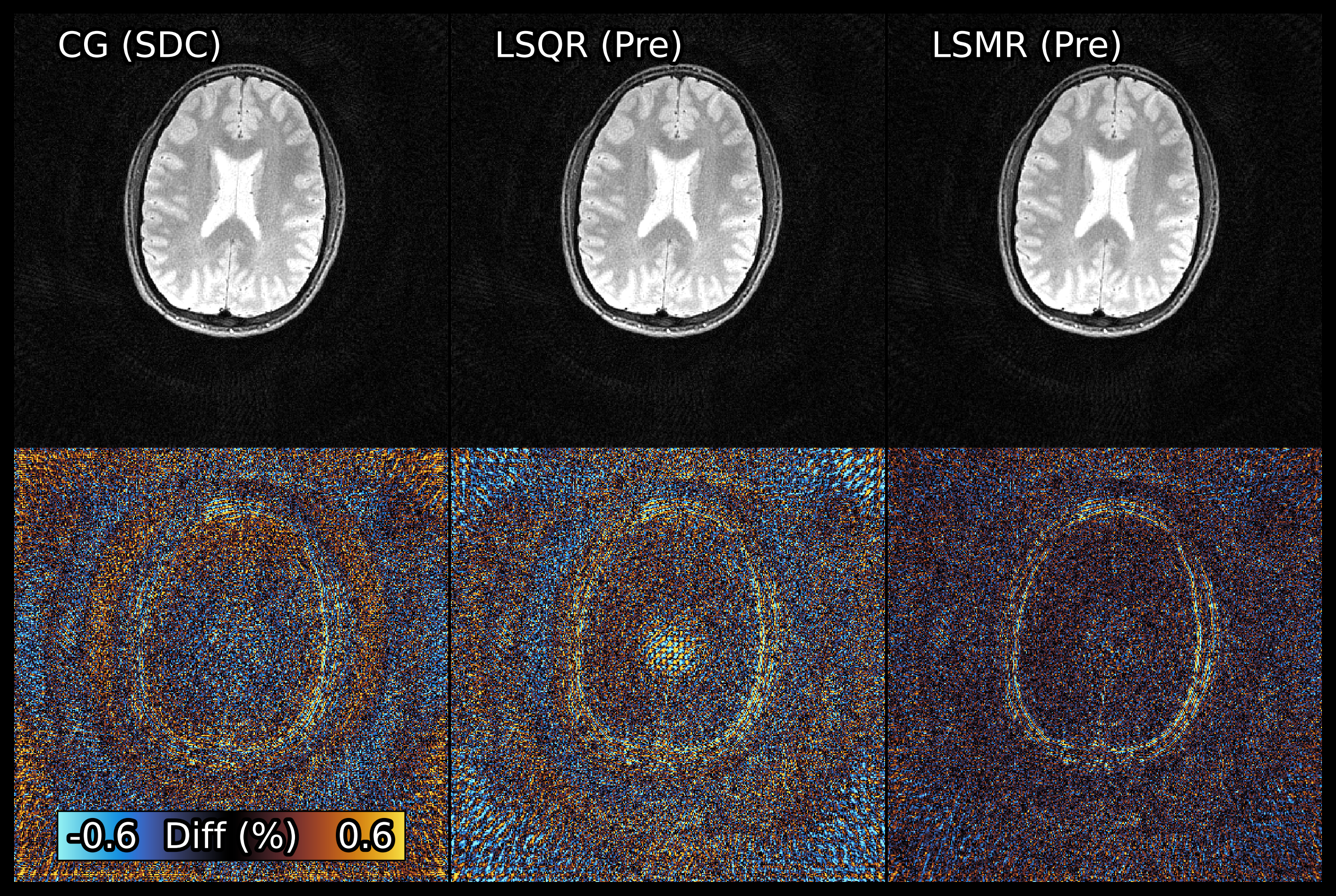}
     \end{subfigure}%
\caption{Reconstructed images at iteration 6 for the 2D data without (left) and with (right) SDC or preconditioning. The bottom row shows the difference from the previous iteration.}
\label{2d-early}
\end{center}
\end{figure*}

\begin{figure*}[htb]
\begin{center}
     \begin{subfigure}[b]{0.5\textwidth}
         \includegraphics[width=\textwidth]{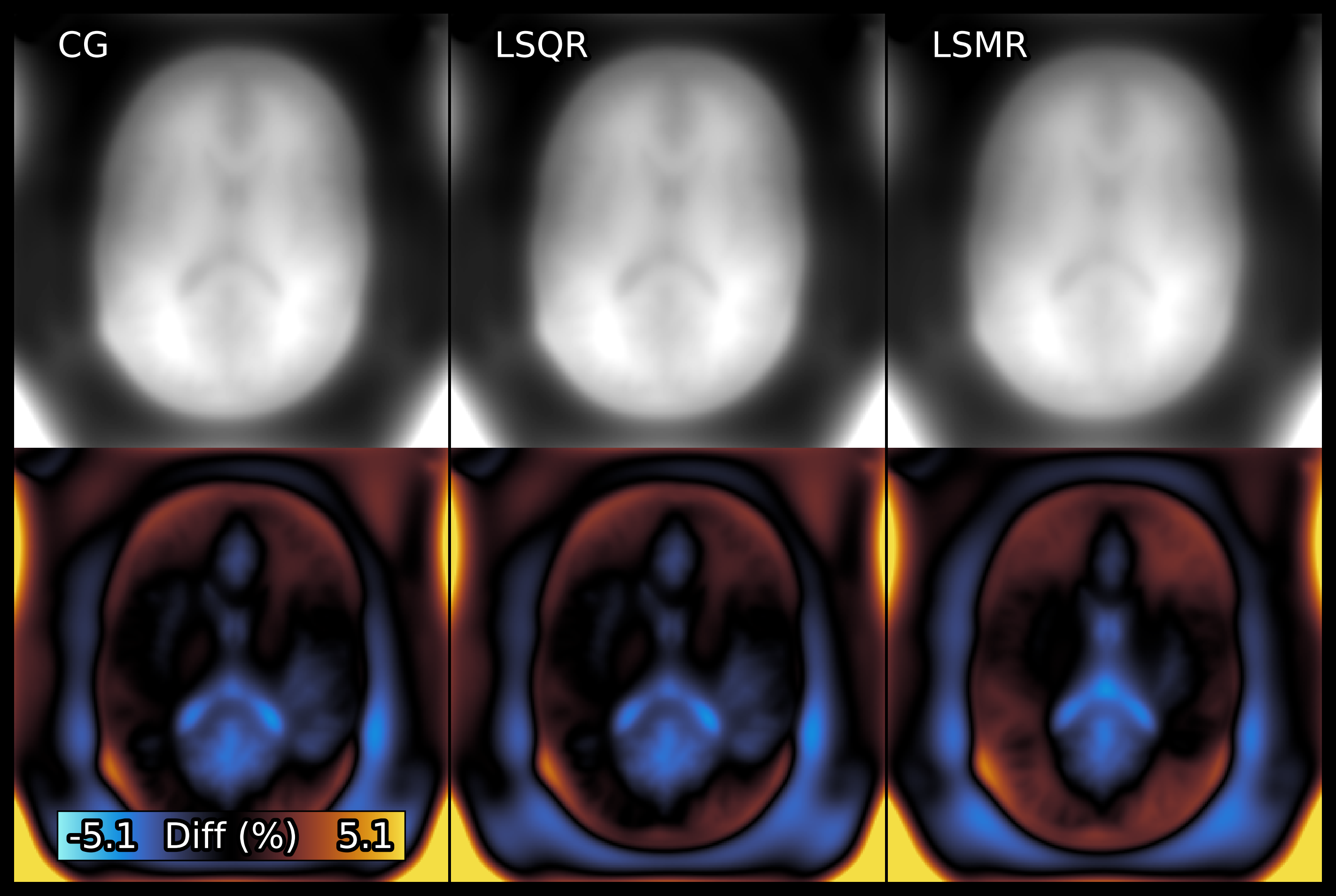}
     \end{subfigure}%
     \begin{subfigure}[b]{0.5\textwidth}
         \includegraphics[width=\textwidth]{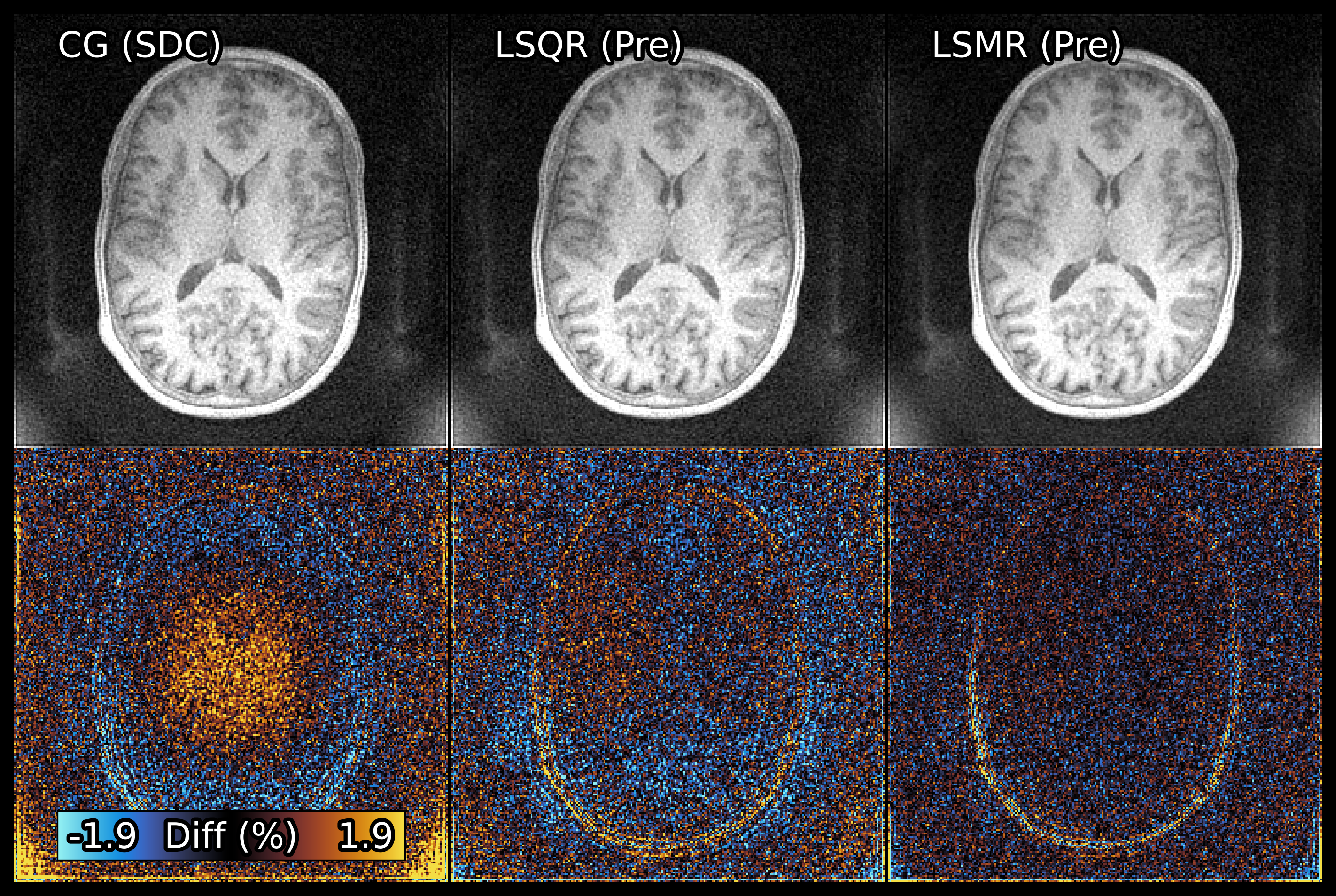}
     \end{subfigure}%
\caption{Reconstructed images iteration 4 for the 3D data without (left) and with (right) SDC or preconditioning. The bottom row shows the difference from the previous iteration.}
\label{3d-early}
\end{center}
\end{figure*}

Figures 5\&6 show the reconstructed images when terminated early, along with the difference from the previous iteration. In the 2D case, without SDC/preconditioning the algorithm has clearly not finished, with significant anatomy present in the change from the previous iteration. With SDC/preconditioning anatomical features are largely absent from the update image but for both CG and LSQR low spatial-frequency changes are obvious. In contrast, with the exception of the edges of the skull, LSMR shows a much flatter update image.

In 3D this pattern is repeated, but more pronounced with the images lacking SDC/preconditioning being exceedingly blurry. With SDC/preconditioning, the update from the previous iteration is again flattest for LSMR, and overall the image quality is significantly better than when the algorithm was run for 32 iterations.

\section*{Discussion}

The results above demonstrate the superior numerical performance of the LSMR algorithm for noncartesian MR reconstruction. LSMR produced the lowest residual values, but more importantly demonstrates monotonic convergence behaviour that is safe to terminate earlier than either CG or LSQR. When preconditioning is used, the image quality after only a handful of iterations is equivalent (2D) or superior (3D) to that without preconditioning after many iterations. There is essentially no computational drawback to using LSMR compared to CG. The majority of algorithmic time is spent in the forward and adjoint Non-Uniform Fast Fourier Transform (NUFFT) operator and is the same for both (when Toeplitz embedding is not used). LSMR does require a larger number of subsequent scalar calculations to find the update step-size, and is hence more tedious to implement, but these calculations take negligible run time.

The question of whether a Toeplitz-embedded reconstruction using CG would still be advantageous over LSMR is one I do not attempt to answer here. Such a comparison would involve numerous trade-offs such as memory consumption versus run-time performance and reconstruction accuracy that will depend on the specifics of the particular MR reconstruction problem. I included the 2D data, which could easily be used with Toeplitz embedding, in this paper to provide an easy reference point to the existing literature, but my particular interest is the 3D ZTE data where for reasons outlined above Toeplitz embedding is not always possible.

I found a limited number of references to using LSQR in the MRI literature~\cite{Hoge:2006,Qu:2007,Haldar:2015}. Of note, Hoge et al exploited the way Tikhonov regularization works in LSQR to provide automatic tuning of the regularization parameter~\cite{Hoge:2006}. LSMR has previously been used to improve image quality for Cartesian GRAPPA~\cite{Weller:2013} and was recently used as the comparison method against a Total-Variation regularized reconstruction of PROPELLER data~\cite{Chen:2021}.

I am not aware of an investigation of the algorithms for general noncartesian imaging as presented here. I hypothesise this is likely due to the fact that an efficient preconditioner for noncartesian reconstruction, which is essential for both efficiency and image quality, was only discovered recently~\cite{Ong:2020}. The single-channel preconditioner as used here can be precomputed for a particular trajectory, as can SDC weightings, and during each iteration only requires a single element-wise multiplication operation in k-space which takes minimal time.

While slightly beyond the scope of this paper, it is worth mentioning that LSMR (or LSQR) can be used as the inner-solver for the Alternating Directions Method-of-Multipliers (ADMM) algorithm which has become popular for MRI reconstruction~\cite{BART, Tamir:2017, Asslander:2018}. An example of how to formulate this with LSQR can be found on the website associated with Boyd et al~\cite{BoydADMM}.

\section*{Conclusion}

In situations where Toeplitz embedding is impractical the LSMR algorithm has numerical benefits over the Conjugate Gradient algorithm and negligible drawbacks.

\section*{Acknowledgements}

I thank Emil Ljungberg for his encouragement and reading a draft of this manuscript.

\bibliographystyle{unsrt}
\bibliography{references}

\begin{thebibliography}{10}

\bibitem{Pruessmann:2001}
Klaas~P. Pruessmann, Markus Weiger, Peter Börnert, and Peter Boesiger.
\newblock Advances in sensitivity encoding with arbitrary k-space trajectories.
\newblock {\em Magnetic Resonance in Medicine}, 46(4):638--651, October 2001.

\bibitem{Maier:2020}
Oliver Maier, Steven~Hubert Baete, Alexander Fyrdahl, Kerstin Hammernik, Seb
  Harrevelt, Lars Kasper, Agah Karakuzu, Michael Loecher, Franz Patzig,
  Ye~Tian, Ke~Wang, Daniel Gallichan, Martin Uecker, and Florian Knoll.
\newblock {CG}‐{SENSE} revisited: {Results} from the first {ISMRM}
  reproducibility challenge.
\newblock {\em Magnetic Resonance in Medicine}, November 2020.

\bibitem{Wajer:2001}
FTAW Wajer and KP~Pruessmann.
\newblock Major speedup of reconstruction for sensitivity encoding with
  arbitrary trajectories.
\newblock In {\em Proc. {Intl}. {Soc}. {Mag}. {Res}. {Med}}, page 767, 2001.

\bibitem{Fessler:2005}
J.A. Fessler, {Sangwoo Lee}, V.T. Olafsson, H.R. Shi, and D.C. Noll.
\newblock Toeplitz-based iterative image reconstruction for {MRI} with
  correction for magnetic field inhomogeneity.
\newblock {\em IEEE Transactions on Signal Processing}, 53(9):3393--3402,
  September 2005.

\bibitem{Baron:2018}
Corey~A. Baron, Nicholas Dwork, John~M. Pauly, and Dwight~G. Nishimura.
\newblock Rapid compressed sensing reconstruction of {3D} non-{Cartesian}
  {MRI}.
\newblock {\em Magnetic Resonance in Medicine}, 79(5):2685--2692, May 2018.

\bibitem{Ong:2020}
Frank Ong, Martin Uecker, and Michael Lustig.
\newblock Accelerating {Non}-{Cartesian} {MRI} {Reconstruction} {Convergence}
  {Using} k-{Space} {Preconditioning}.
\newblock {\em IEEE Transactions on Medical Imaging}, 39(5):1646--1654, May
  2020.

\bibitem{Pipe:2000}
James~G. Pipe.
\newblock Reconstructing {MR} images from undersampled data: {Data}-weighting
  considerations.
\newblock {\em Magnetic Resonance in Medicine}, 43(6):867--875, June 2000.

\bibitem{Johnson:2009}
Kenneth~O. Johnson and James~G. Pipe.
\newblock Convolution kernel design and efficient algorithm for sampling
  density correction.
\newblock {\em Magnetic Resonance in Medicine}, 61(2):439--447, February 2009.

\bibitem{Zwart:2012}
Nicholas~R. Zwart, Kenneth~O. Johnson, and James~G. Pipe.
\newblock Efficient sample density estimation by combining gridding and an
  optimized kernel: {Efficient} {Sample} {Density} {Estimation}.
\newblock {\em Magnetic Resonance in Medicine}, 67(3):701--710, March 2012.

\bibitem{Ljungberg:2021}
Emil Ljungberg, Nikou Damestani, Tobias~C Wood, David~J Lythgoe, Fernando
  Zelaya, Steven C~R Williams, Ana~Beatriz Solana, Gareth~J Barker, and Florian
  Wiesinger.
\newblock Silent zero {TE} {MR} neuroimaging: {Current} state-of-the-art and
  future directions.
\newblock {\em Progress in Nuclear Magnetic Resonance Spectroscopy}, page~21,
  2021.

\bibitem{Weiger:2019}
Markus Weiger and Klaas~P. Pruessmann.
\newblock Short-{T2} {MRI}: {Principles} and recent advances.
\newblock {\em Progress in Nuclear Magnetic Resonance Spectroscopy},
  114-115:237--270, October 2019.

\bibitem{Weiger:2020}
Markus Weiger, Romain Froidevaux, Emily~Louise Baadsvik, David~Otto Brunner,
  Manuela~Barbara Rösler, and Klaas~Paul Pruessmann.
\newblock Advances in {MRI} of the myelin bilayer.
\newblock {\em NeuroImage}, 217:116888, August 2020.

\bibitem{Baadsvik:2021}
Emily~Louise Baadsvik, Markus Weiger, Romain Froidevaux, Manuela~Barbara
  Rösler, David~Otto Brunner, Lena Öhrström, Frank~Jakobus Rühli, Patrick
  Eppenberger, and Klaas~Paul Pruessmann.
\newblock High‐resolution {MRI} of mummified tissues using advanced
  short‐{T} $_{\textrm{2}}$ methodology and hardware.
\newblock {\em Magnetic Resonance in Medicine}, 85(3):1481--1492, March 2021.

\bibitem{LSQR}
Christopher~C. Paige and Michael~A. Saunders.
\newblock {LSQR}: {An} {Algorithm} for {Sparse} {Linear} {Equations} and
  {Sparse} {Least} {Squares}.
\newblock {\em ACM Transactions on Mathematical Software}, 8(1):43--71, March
  1982.

\bibitem{LSMR}
David Chin-Lung Fong and Michael Saunders.
\newblock {LSMR}: {An} {Iterative} {Algorithm} for {Sparse} {Least}-{Squares}
  {Problems}.
\newblock {\em SIAM Journal on Scientific Computing}, 33(5):2950--2971, January
  2011.

\bibitem{Orban:2020}
A.~Montoison, D.~Orban, and {contributors}.
\newblock Krylov.jl: {A} {Julia} basket of hand-picked {Krylov} methods, June
  2020.

\bibitem{Emil:2022}
Emil Ljungberg, Tobias~C. Wood, Ana~Beatriz Solana, Steven C.~R. Williams,
  Gareth~J. Barker, and Florian Wiesinger.
\newblock Motion corrected silent {ZTE} neuroimaging.
\newblock {\em Magnetic Resonance in Medicine}, page mrm.29201, April 2022.

\bibitem{Wood:2021}
Tobias Wood, Emil Ljungberg, and Florian Wiesinger.
\newblock Radial {Interstices} {Enable} {Speedy} {Low}-volume {Imaging}.
\newblock {\em Journal of Open Source Software}, 6(66):3500, October 2021.

\bibitem{Huang:2008}
Feng Huang, Sathya Vijayakumar, Yu~Li, Sarah Hertel, and George~R. Duensing.
\newblock A software channel compression technique for faster reconstruction
  with many channels.
\newblock {\em Magnetic Resonance Imaging}, 26(1):133--141, January 2008.

\bibitem{Yeh:2005}
Ernest~N. Yeh, Matthias Stuber, Charles~A. McKenzie, Rene~M. Botnar, Tim
  Leiner, Michael~A. Ohliger, Aaron~K. Grant, Jacob~D. Willig-Onwuachi, and
  Daniel~K. Sodickson.
\newblock Inherently self-calibrating non-cartesian parallel imaging.
\newblock {\em Magnetic Resonance in Medicine}, 54(1):1--8, July 2005.

\bibitem{Hoge:2006}
W.S. Hoge, M.E. Kilmer, S.J. Haker, D.H. Brooks, and W.E. Kyriakos.
\newblock Fast {Regularized} {Reconstruction} of {Non}-{Uniformly} {Subsampled}
  {Parallel} {MRI} {Data}.
\newblock In {\em 3rd {IEEE} {International} {Symposium} on {Biomedical}
  {Imaging}: {Macro} to {Nano}, 2006.}, pages 714--717, Arlington, Virginia,
  USA, 2006. IEEE.

\bibitem{Qu:2007}
Peng Qu, Jing Luo, Bida Zhang, Jianmin Wang, and Gary~X. Shen.
\newblock An improved iterative {SENSE} reconstruction method.
\newblock {\em Concepts in Magnetic Resonance Part B: Magnetic Resonance
  Engineering}, 31B(1):44--50, February 2007.

\bibitem{Haldar:2015}
Justin~P. Haldar.
\newblock Autocalibrated loraks for fast constrained {MRI} reconstruction.
\newblock In {\em 2015 {IEEE} 12th {International} {Symposium} on {Biomedical}
  {Imaging} ({ISBI})}, pages 910--913, Brooklyn, NY, USA, April 2015. IEEE.

\bibitem{Weller:2013}
Daniel~S. Weller, Jonathan~R. Polimeni, Leo Grady, Lawrence~L. Wald, Elfar
  Adalsteinsson, and Vivek~K. Goyal.
\newblock Sparsity-{Promoting} {Calibration} for {GRAPPA} {Accelerated}
  {Parallel} {MRI} {Reconstruction}.
\newblock {\em IEEE Transactions on Medical Imaging}, 32(7):1325--1335, July
  2013.

\bibitem{Chen:2021}
Hsin-Chia Chen, Haw-Chiao Yang, Chih-Ching Chen, Seb Harrevelt, Yu-Chieh Chao,
  Jyh-Miin Lin, Wei-Hsuan Yu, Hing-Chiu Chang, Chin-Kuo Chang, and Feng-Nan
  Hwang.
\newblock Improved {Image} {Quality} for {Static} {BLADE} {Magnetic}
  {Resonance} {Imaging} {Using} the {Total}-{Variation} {Regularized} {Least}
  {Absolute} {Deviation} {Solver}.
\newblock {\em Tomography}, 7(4):555--572, October 2021.

\bibitem{BART}
Martin Uecker, Frank Ong, Jonathan~I Tamir, Dara Bahri, Patrick Virtue,
  Joseph~Y Cheng, Tao Zhang, and Michael Lustig.
\newblock Berkeley advanced reconstruction toolbox.
\newblock In {\em Proc. {Intl}. {Soc}. {Mag}. {Reson}. {Med}}, volume~23, 2015.

\bibitem{Tamir:2017}
Jonathan~I. Tamir, Martin Uecker, Weitian Chen, Peng Lai, Marcus~T. Alley,
  Shreyas~S. Vasanawala, and Michael Lustig.
\newblock T2 shuffling: {Sharp}, multicontrast, volumetric fast spin‐echo
  imaging.
\newblock {\em Magnetic Resonance in Medicine}, 77(1):180--195, January 2017.

\bibitem{Asslander:2018}
Jakob Assländer, Martijn~A. Cloos, Florian Knoll, Daniel~K. Sodickson, Jürgen
  Hennig, and Riccardo Lattanzi.
\newblock Low rank alternating direction method of multipliers reconstruction
  for {MR} fingerprinting: {Low} {Rank} {ADMM} {Reconstruction}.
\newblock {\em Magnetic Resonance in Medicine}, 79(1):83--96, January 2018.

\bibitem{BoydADMM}
Stephen Boyd.
\newblock Distributed {Optimization} and {Statistical} {Learning} via the
  {Alternating} {Direction} {Method} of {Multipliers}.
\newblock {\em Foundations and Trends® in Machine Learning}, 3(1):1--122,
  2010.

\end{thebibliography}
\end{document}